\documentstyle[twocolumn,aps,prl,epsf,floats]{revtex}
\begin{document}
\draft
\flushbottom
\twocolumn[
\hsize\textwidth\columnwidth\hsize\csname @twocolumnfalse\endcsname

\title{Slave Boson Approach to The Neutron Scattering in
       YBa$_2$Cu$_3$O$_{6+y}$ Superconductors}

\author{Jan Brinckmann\cite{tkmaddr} and Patrick A.\ Lee}

\address{Dept.\ of Physics, Massachusetts Institute of Technology, Cambridge,
         Massachusetts 02139}

\widetext

\date{ 3 Nov.\ 1998\,, Vers.\ 2 }
\maketitle
\tightenlines
\widetext
\advance\leftskip by 57pt
\advance\rightskip by 57pt

\begin{abstract}
%
The evolution of the so-called ``41\,meV resonance'' in the magnetic
response of YBCO cuprates is studied with slave-boson theory for the
$t$--$t'$--$J$-model. The resonance appears as a collective
spin fluctuation in the d-wave superconducting (SC) state. It is
undamped at optimal doping due to a
threshold in the excitation energies of particle--hole pairs
with relative wave vector $(\pi,\pi)$\,. When hole filling is reduced,
the resonance moves to lower energies and broadens. Below the
resonance energy we find a crossover to an incommensurate response in
agreement with a recent experiment on YBa$_2$Cu$_3$O$_{6.6}$\,.
We show that dynamic nesting in the d-wave SC state causes this effect.
%
%
\end{abstract}

]
\narrowtext

\newlength{\mysize}
\def\loadepsfig#1{
 \def\figname{#1}
 \vbox to 10pt {\ }
 \vbox{ \hbox to \hsize {
   \mysize\hsize    \advance \mysize by -20pt
   \def\epsfsize##1##2{\ifdim##1>\mysize\mysize\else##1\fi}
   \hfill \epsffile{\figname.eps} \hfill
        } }
 \vbox to 7pt {\ }
 }
\def\gloscale{0.8}
\def\dispfig#1{
  \def\figname{#1}
  \def\epsfsize##1##2{\gloscale##1}
  \settowidth{\mysize}{ \epsffile{\figname.eps} }
  \parbox{\mysize}{ \epsffile{\figname.eps} }
  }
\newlength \mygapsize

One of the puzzling phenomena in copper-oxide
superconductors is the so-called ``41\,meV resonance'' observed in
inelastic neutron-scattering (INS) experiments on
optimally doped Y\,Ba$_2$Cu$_3$O$_{6 + y}$ (YBCO)
compounds in the superconducting (SC) phase
\cite{ros91,tra92,moo93,fon95,bou96}\,.
The magnetic susceptibility $\chi''({\bf q},\omega)$ shows a peak at the
antiferromagnetic (AF) wave vector $(\pi,\pi)$
and an energy $\omega_0=$41\,meV\,.
The peak appears to be resolution limited, i.e., almost undamped, and
vanishes above $T_c$\,. In underdoped YBCO the peak shifts to lower
energies $\omega_0 < 41$\,meV \cite{dai96,fon97,bou97} and
develops some damping. It is also visible in the spin-gap regime above
$T_c$\,.
Several theoretical approaches to the magnetic excitations have been
proposed
\cite{tan94,liu95,ste94,bul96,takmor98pre,dahm98pre,dem95}\,.
In most cases \cite{tan94,liu95,ste94,bul96,takmor98pre,dahm98pre}
the resonance at optimal doping is
identified as a collective spin fluctuation, which is stabilized
through the suppression of quasi-particle damping in the SC state, and
vanishes in the normal phase.
The question arises how this mode evolves with decreasing doping.
In particular it is unclear how the finite damping
at intermediate doping levels can be obtained, since the peak is
expected to {\em narrow} and become a spin-wave (Bragg) peak at zero
energy when the N{\'e}el state is reached.
There is also experimental evidence \cite{tra92,mook98}
that the magnetic response in YBCO at energies somewhat below
$\omega_0$ is incommensurate. It is dominated by four
peaks at ${\bf q}=(\pi, \pi\pm\delta)$ and ${\bf q}=(\pi\pm\delta,
\pi)$\,, a pattern previously known only for the
La$_{2-x}$Sr$_x$CuO$_{4+y}$ family of compounds.
In view of the new experimental data, we re-investigate the
spin response in slave-boson mean-field theory
\cite{tan94,liu95,ste94}\,, focusing on
the evolution of the resonance with hole filling (doping) and the
crossover commensurate--incommensurate with variation of energy.

We start from the t--t'--J-model for a single CuO$_2$-layer.
Results specific to the bi-layer structure of YBCO will be presented
elsewhere. The model reads
\begin{eqnarray*}
  H & = &
    - \sum_{(i,j), \sigma} t \,c^\dagger_{i \sigma} c_{j \sigma}
    - \sum_{(i,j)', \sigma} t' \,c^\dagger_{i \sigma} c_{j \sigma}
    + \frac{1}{2}\sum_{(i,j)} J \,{\bf S}_i {\bf S}_j
\end{eqnarray*}
Sums include nearest neighbor $(i,j)$ or next n.n.\ $(i,j)'$
Cu-sites on a 2D square lattice.
Physical operators $c_{i \sigma}$ = $b^\dagger_i f_{i \sigma}$\,,
${\bf S}_i$ = $\frac{1}{2}\sum_{\sigma, \sigma'}$ $f^\dagger_{i \sigma}
\bbox{\tau}^{\sigma \sigma'} f_{i \sigma'}$
are represented by auxiliary (`slave') bosons $b_i$ carrying the
charge and fermions $f_{i\sigma}$ representing the spin $\sigma$\,.
In mean-field theory the d-wave
superconducting phase is represented by condensed bosons
$b_i \to \langle b_i \rangle$
and fermions in a d-wave BCS state with dispersion
\begin{displaymath}
  \varepsilon({\bf k}) =
    -2 \widetilde{t}[ \cos(k_x) + \cos(k_y)]
    - 4 \widetilde{ t}'\cos(k_x) \cos(k_y) - \mu_f
\end{displaymath}
and gap
$\Delta({\bf k}) =
    \Delta_0 [ \cos(k_x) - \cos(k_y)] / 2$\,.
Renormalized hopping parameters are
$\widetilde{ t} =
    x t + \case{1}{2}J\langle f^\dagger_{i\uparrow}
          f_{i + \hat{x} \uparrow} \rangle$\,,
$\widetilde{ t}' =
    x t'$\,.
The expectation values $\langle \ldots \rangle$ in $\widetilde{ t}$ and
$\Delta_0 =
    2 J [ \langle f_{i\uparrow} f_{i + \hat{x}\downarrow} \rangle
          - \langle f_{i\downarrow} f_{i + \hat{x}\uparrow} \rangle ]$
are computed self-consistently from minimization of the free energy.
The hole concentration (doping) $x$ sets the density of fermions
$(1 - x) =
    \sum_\sigma\langle f^\dagger_{i \sigma} f_{i \sigma} \rangle$
and bosons
$\langle b_i \rangle = \sqrt{x}$\,.
\begin{figure}[htb]
\def\gloscale{0.45}
%
\def\saft{\;\,}
\begin{displaymath}
    \chi^0({\bf q},\omega) \saft= \saft
       \dispfig{susbub1} \saft + \saft
       \dispfig{susbub2}
\end{displaymath}
%
 %
%
%
\caption[\ ]{
  Particle--hole (ph) irreducible contribution to the susceptibility Eq.\
  (\ref{eqn-rpa})\,. Single and double arrowed lines are
  normal and anomalous Green's functions of auxiliary fermions in the
  d-wave BCS state. Shaded areas denote the vertex function
  $\Lambda=1+\Lambda^J$\,. $\Lambda^J$ accounts for all ladder diagrams
  not included in a simple random-phase approximation.
  }
 \label{fig-bubble}
\end{figure}

The magnetic susceptibility is calculated in ladder approximation,
\begin{equation}  \label{eqn-rpa}
  \chi({\bf q},\omega) =
    \chi^0({\bf q},\omega) / [ 1 + \alpha J({\bf q}) \chi^0({\bf q},\omega) ]
\end{equation}
with
$J({\bf q}) =
  J [ \cos(q_x) + \cos(q_y) ]$
and $\alpha = 1$\,.
Fig.\ \ref{fig-bubble} shows $\chi^0({\bf q},\omega)$\,, which is
built of fermion bubbles with a vertex function to include all ladder
diagrams. Boson excitations do
not appear in ladder approximation. Therefore the calculation does
not distinguish between the SC and the spin-gap phase ($\Delta_0\ne 0$
but $\langle b_i \rangle=0$), and the persistence of the resonance
above $T_c$ in underdoped cuprates appears quite naturally.

If $\chi({\bf q},\omega=0)$ is computed from Eq.\ (\ref{eqn-rpa})\,,
an instability to an (incommensurate) N{\'e}el state
occurs for hole densities $x<x_c \approx 0.2$\,. The high value
of $x_c$ is an artifact of the mean-field theory. Since the inclusion
of fluctuations is beyond the scope of this work, we model
the expected suppression \cite{kha93,kyu98pre} by setting
$\alpha=0.34$ in Eq.(\ref{eqn-rpa}), such that $x_c$ is reduced to an
observed value $0.02$\,. It is assumed that the
superconducting phase $x>x_c$ is stable when fluctuations
are included, at least for low $T$\,. A renormalization of the
mean-field parameters, e.g., $\Delta_0$ is ignored, in order to
keep $\alpha\leftrightarrow x_c$ the only phenomenological
input. The self-consistent
$\Delta_0\approx 45$\,meV at optimal doping actually compares to
experimental values \cite{ren98ref}\,. We have checked
numerically that the local moment sum-rule is not sensitive to the
choice of $\alpha$\,.

\begin{figure}[htb]
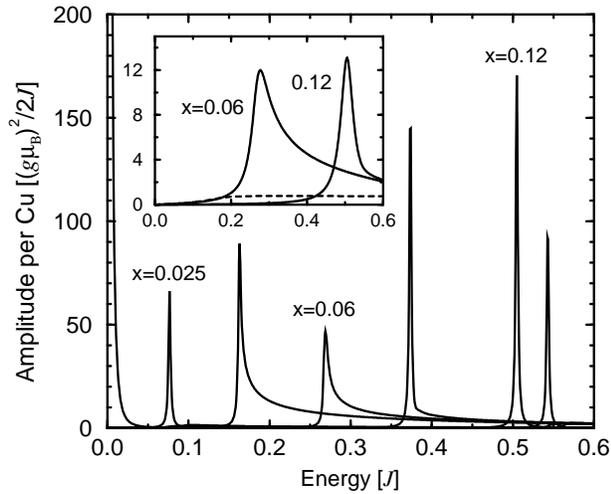

 \loadepsfig{figresnorm}
\caption[\ ]{
  Imaginary part $\chi''$ of the susceptibility Eq.(\ref{eqn-rpa}) for
  wave vector $(\pi, \pi)$\,.
  {\bf Main figure:} superconducting state at $T\to 0$\,;
  different peaks belong from left to right to hole
  concentrations $x=0.02$ (scaled $\times\, 0.1$), $0.025$ ($\times\,
  0.1$), $0.04$, $0.06$, $0.08$, $0.12$, $0.16$\,. All peaks except
  $x=0.04, 0.06, 0.08$ are $\delta$-functions, computed with a
  quasi-particle damping $\Gamma=10^{-3}J$\,.
  {\bf Inset:} $x=0.06, 0.12$ in the SC state (full lines) and
  $x=0.12$ in the normal
  state $T\gtrsim T_c$ (dashed line). Here an experimental energy
  resolution (FWHM) of 5\,meV is simulated via $\Gamma=0.01J$\,.
  }
 \label{fig-reson}
\end{figure}

With this approach we are able to access the full range of hole densities
$0.02<x<0.15$ of underdoped cuprates. Numerical calculations are performed
in the superconducting state at low temperature $T\to 0$ with
parameters $t=2J$\,, $t'=-0.45t$ for YBCO \cite{okand94,schabel98ii}\,.
Fig.\ \ref{fig-reson}
shows the imaginary part $\chi''({\bf Q}_\pi, \omega)$ of
Eq.(\ref{eqn-rpa}) at the AF wave
vector ${\bf Q}_\pi \equiv (\pi,\pi)$ for several hole
concentrations. Apparently $\chi''$ is dominated by a sharp
resonance. For $x=0.12$ it appears at an energy $\omega_0 = 0.51J
\approx 60$\,meV\,, its residual width is due to the small quasi-particle
damping used in the numerical
calculation. When doping is reduced to $x\approx 0.08 \to
0.04$ the resonance moves monotonously to lower energies and
develops some damping. For further reduced hole filling, the peak
becomes again resolution limited and eventually shifts to
$\omega_0 \to 0$ when the AF transition is reached at $x = x_c = 0.02$\,.
The peak vanishes in the normal phase $T>T_c$ near optimal doping,
see the inset in Fig.\ \ref{fig-reson}\,. In heavily
overdoped systems $x > 0.2$ the resonance vanishes even in the SC
state.

The qualitative agreement between these results
and the experimental findings summarized in
the beginning is quite satisfactory. The resonance position
$\omega_0\approx 60$\,meV near optimal doping ($x=0.12)$ is not too
far off the observed value 40\,meV\,. We also calculated
the AF correlation length $\zeta$ from the equal-time correlation function
$\langle {\bf S}({\bf q}) {\bf S}(-{\bf q}) \rangle$\,.
When doping is reduced, $\zeta$ increases
monotonically and reaches the system size at $x =
x_c = 0.02$\,. Quantitatively $\zeta$ is overestimated by
a factor of $\approx 2$ compared to known values \cite{bir88,singlen92}\,.

\begin{figure}[htb]
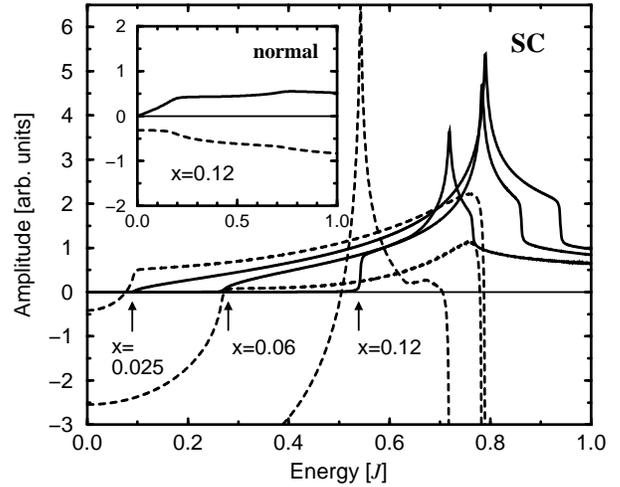

 \loadepsfig{fignnnpoles}
\caption[\ ]{
  Imaginary part $\chi^{0}{}''$ of the
  particle--hole bubbles from Fig.\ \ref{fig-bubble} for wave vector
  $(\pi,\pi)$ (continuous lines), and the inverse Stoner enhancement-factor $K$
  defined in the text (scaled $\times (-10)$\,, dashed lines).
  {\bf Main figure:} shown are three hole
  fillings $x$ in the SC state at $T\to 0$\,. An arrow
  points to the threshold energy $\Omega_0$ in the respective
  $\chi^{0}{}''$\,; the corresponding $K$ crosses zero at
  $\omega_0$ nearby. $\chi^0{}''$ shows a van Hove singularity at
  $2\Delta_0\ge 0.7J$ (a peak) and for $x=0.12$ also at
  $\Omega_0\approx 0.53J$ (a step).
  {\bf Inset:} normal state $T\gtrsim T_c$ for $x=0.12$\,.
  }
 \label{fig-poles}
\end{figure}

It is at first sight surprising that the resonance occurs
at energies $\omega_0 \lesssim 2\Delta_0$ without significant damping
from particle--hole (ph) excitations, since the d-wave SC phase has a
finite density of states. Fig.\
\ref{fig-poles} shows the imaginary part $\chi^0{}''({\bf
Q}_\pi,\omega)$ of the
ph-bubbles $\chi^0 = \chi^0{}' + i\chi^0{}''$ from
Fig.\ \ref{fig-bubble}\,, together with the real part of the
denominator of Eq.(\ref{eqn-rpa})\,,
$K({\bf Q}_\pi,\omega) =
  [ 1 + \alpha J({\bf Q}_\pi) \chi^0{}'({\bf Q}_\pi,\omega)]$\,.
From the numerical calculation the vertex corrections have no effect on
the outcome of Eq.(\ref{eqn-rpa}) and are omitted here.
$\chi^0{}''$ has a full gap up to an energy $\Omega_0$\,, which
increases with doping. A pole occurs in Eq.(\ref{eqn-rpa}) since the
corresponding $K$ crosses zero at an energy $\omega_0 < \Omega_0$
in the gap. The result is a $\delta$-like resonance for a hole
filling near the AF transition ($x=0.025$) or near optimal doping
($x=0.12$)\,. In
the underdoped case ($x=0.06$) we have $\omega_0\approx\Omega_0$\,,
and the resonance is asymmetrically broadened.
To explain this we note that apart from BCS coherence factors,
%
$\chi^0{}''({\bf q},\omega) \sim
    \sum_{k} \delta(\omega - \Omega({\bf q},{\bf k}))$
at $T\to 0$\,, where
$\Omega({\bf q}, {\bf k}) =
  E({\bf k}) + E({\bf k} + {\bf q})$
and
$E({\bf k}) =
  \sqrt{\varepsilon^2({\bf k}) + \Delta^2({\bf k})}$\,.
Here $\Omega({\bf q}, {\bf k})$ denotes the ph-excitation energies of
fermions with relative wave vector ${\bf q}$\,.
%
%
For ${\bf q}={\bf Q}_\pi$ this has a minimum value $\Omega_0$\,, which
determines the threshold in $\chi^0{}''$\,. It is given by
$\Omega_0=2|\mu_f| Z$\,, where $\mu_f$ is the fermion
chemical potential and $Z=1$
for $\kappa=\Delta_0^2/(4\mu_f\widetilde{t}')>2$ and
$Z=\sqrt{\kappa-\kappa^2/4}$ for $0<\kappa<2$\,.
Note that $\Omega_0$\,, and therefore the resonance energy
$\omega_0$ do not follow $2\Delta_0$\,.

The reason for the absence of damping at optimal
doping ($x=0.12$) is identified in the step-like van Hove singularity
(v.H.s.)\ at $\omega=\Omega_0$ in $\chi^0{}''$ (see Fig.\
\ref{fig-poles}), induced by a
locally flat ph-dispersion. By
virtue of the Kramers--Kroenig transformation the real part $\chi^0{}'$
develops a sharp structure around $\Omega_0$\,,
shifting the position $\omega_0$ of the resonance (i.e., the zero
crossing of $K$) well into the gap. When $x$ is reduced,
the v.H.s.\ and with it the peak structure in $\chi^0{}'$ weaken, and
$\omega_0$ moves close to and may cross the threshold $\Omega_0$ to
damping ph-excitations. At very low doping $x\gtrsim x_c$ this trend is
over-compensated by the monotonous increase of $\chi^0$ with reduced
$x$\,, which shifts $\omega_0$
back into the gap. The increase of $\chi^0$ comes from the shrinking of
the upper cutoff $2\widetilde{ W}$ for ph-excitations, with the bandwidth
$\widetilde{ W}\approx 8\widetilde{t} \approx 8(x t + J/8)$
of Gutzwiller renormalized fermions.
It should be noted that the step-like v.H.s.\
in $\chi^0{}''$ near optimal doping depends on the coexistence of a
finite $\Delta_0$ and effective n.n.n.\ hopping $\widetilde{ t}'$\,,
which lead to a sufficiently flat
$\Omega({\bf Q}_\pi,{\bf k})$\,. Setting $t'=0$ eliminates
the step, and the resonance from Eq.(\ref{eqn-rpa}) is severely broadened.
In the normal phase $\Delta_0=0$ no resonance appears at all. The
ph-dispersion
$\Omega({\bf Q}_\pi, {\bf k}) =
   |\varepsilon({\bf k})| + |\varepsilon({\bf k} + {\bf Q}_\pi)|$
then has a zero minimum value without v.H.s., resulting in a
gapless and structureless $\chi^0{}''$ and $K$ (see the inset of
Fig.\ \ref{fig-poles}).

The approach presented here supports the understanding of the ``41\,meV
resonance'' as a collective spin fluctuation. We find that the sharp
resonance is entirely caused by the pole in the random-phase
approximation (RPA), which sums up spin-singlet
particle--hole (ph) excitations, i.e., transversal
spin fluctuations.  The vertex function
$\Lambda=1 + \Lambda^J$ in $\chi^0$ (see Fig.\ \ref{fig-bubble}) has
numerically no effect ($\Lambda^J \approx 0$) in the energy and
doping range considered here. That is, we can neglect in particular the resonant
contribution from spin-triplet particle--particle (pp) pairs,
which are involved through the mixing with ph-excitations
$f^\dagger_{\bbox{k} + \bbox{Q}_\pi \uparrow}
   f_{\bbox{k} \downarrow} \leftrightarrow
   f^\dagger_{\bbox{k} + \bbox{Q}_\pi \uparrow}
   f^\dagger_{-\bbox{k} \uparrow}$
in the SC state \cite{dem95}\,. Our view on the
neutron scattering differs from, e.g., Ref.\
\cite{dem95}\,, where the pp-channel is considered the
main contribution to the ``41\,meV resonance''. Some further comparison
of these viewpoints has been given in Refs.\ \cite{bri98,eder98}\,.

\begin{figure}[htb]
  \mysize 0.72\hsize
  \mygapsize 0.05\hsize
\begin{center}
  \def\epsfsize#1#2{\epsfxsize}
  \epsfxsize=\mysize
  \parbox{\mysize}{ \epsffile{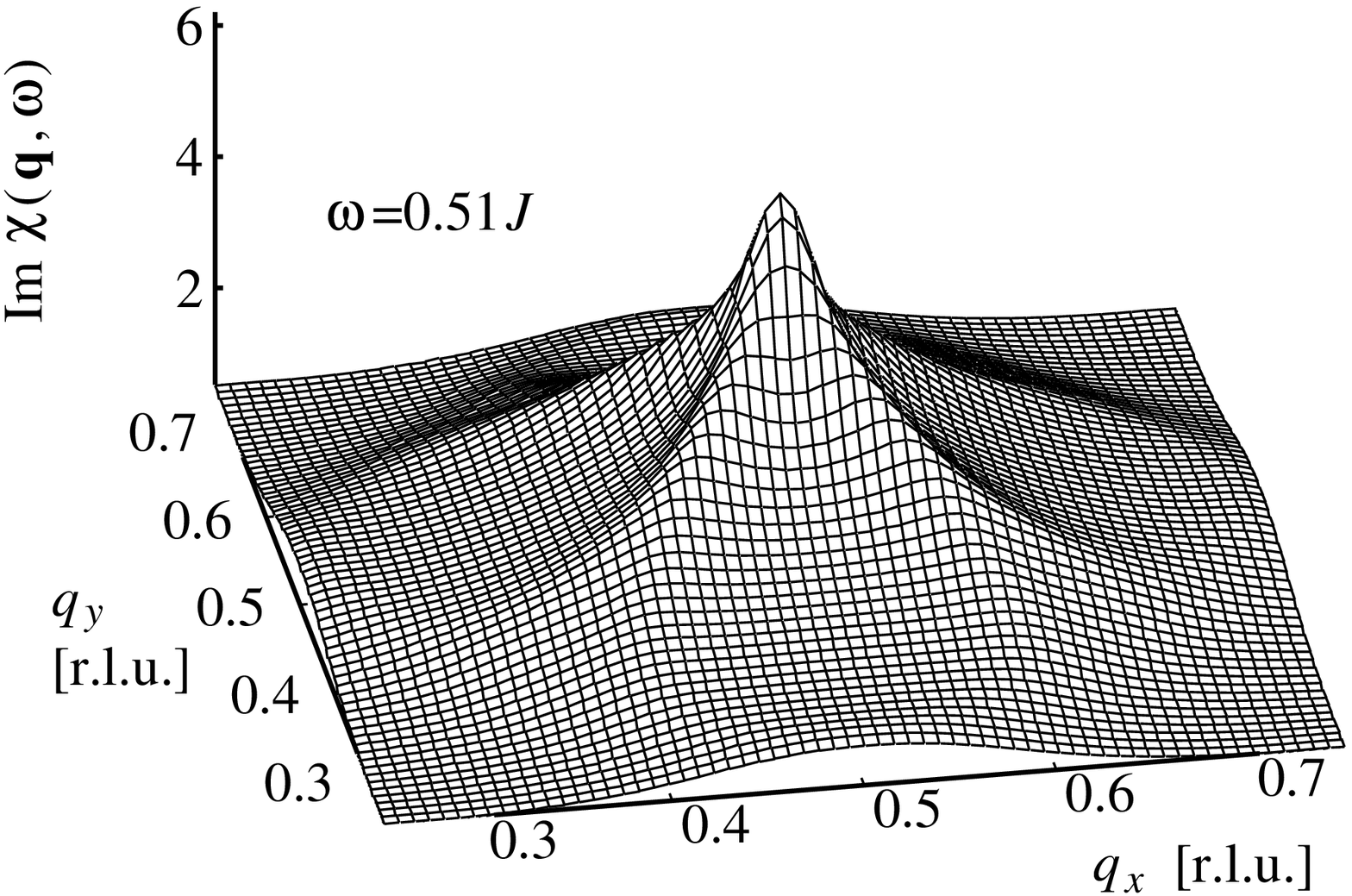} } \\
  \vspace*{-\mygapsize}
  \epsfxsize=\mysize
  \parbox{\mysize}{ \epsffile{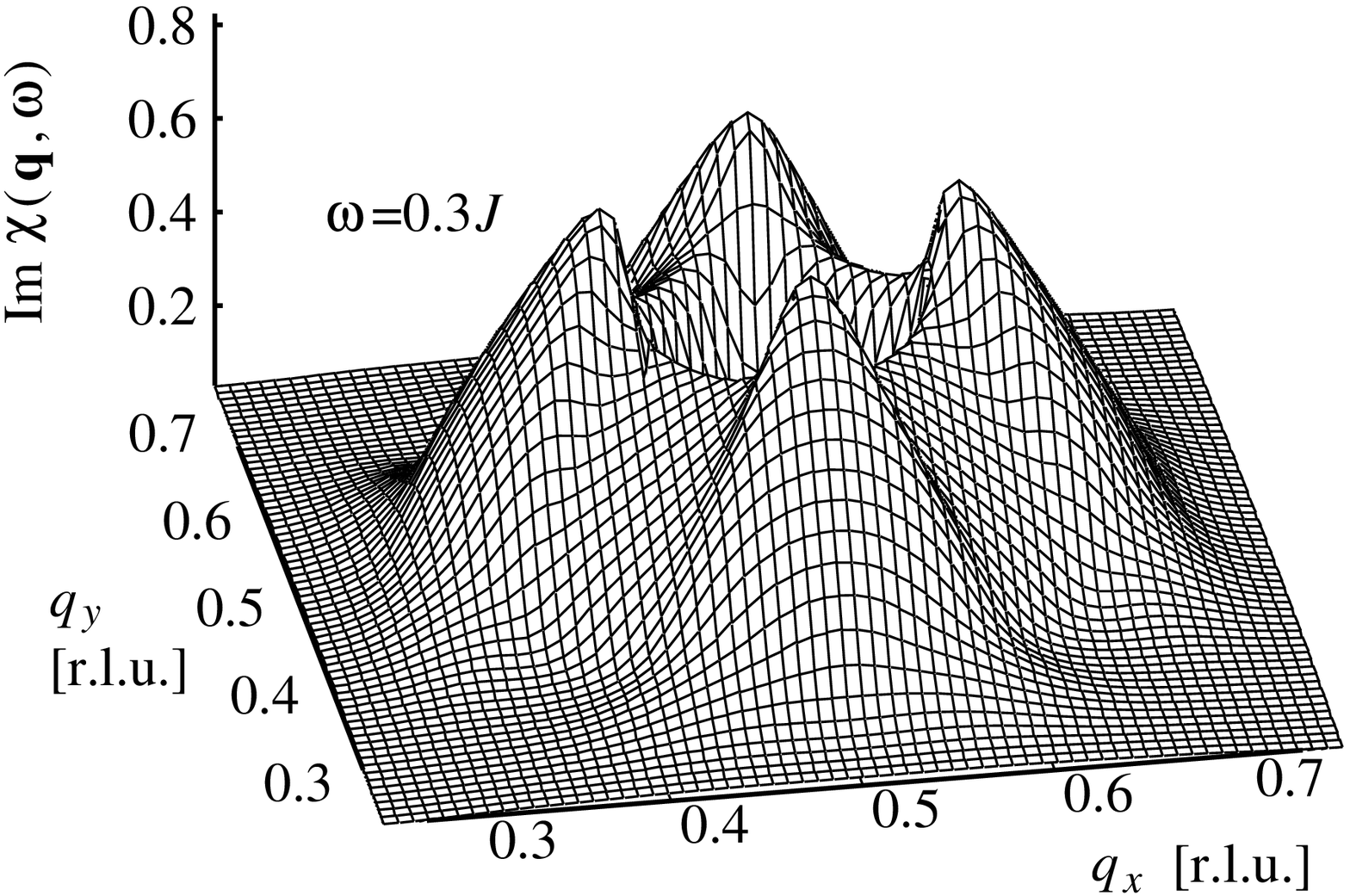} } \\
  \vspace*{-\mygapsize}
  \epsfxsize=\mysize
  \parbox{\mysize}{ \epsffile{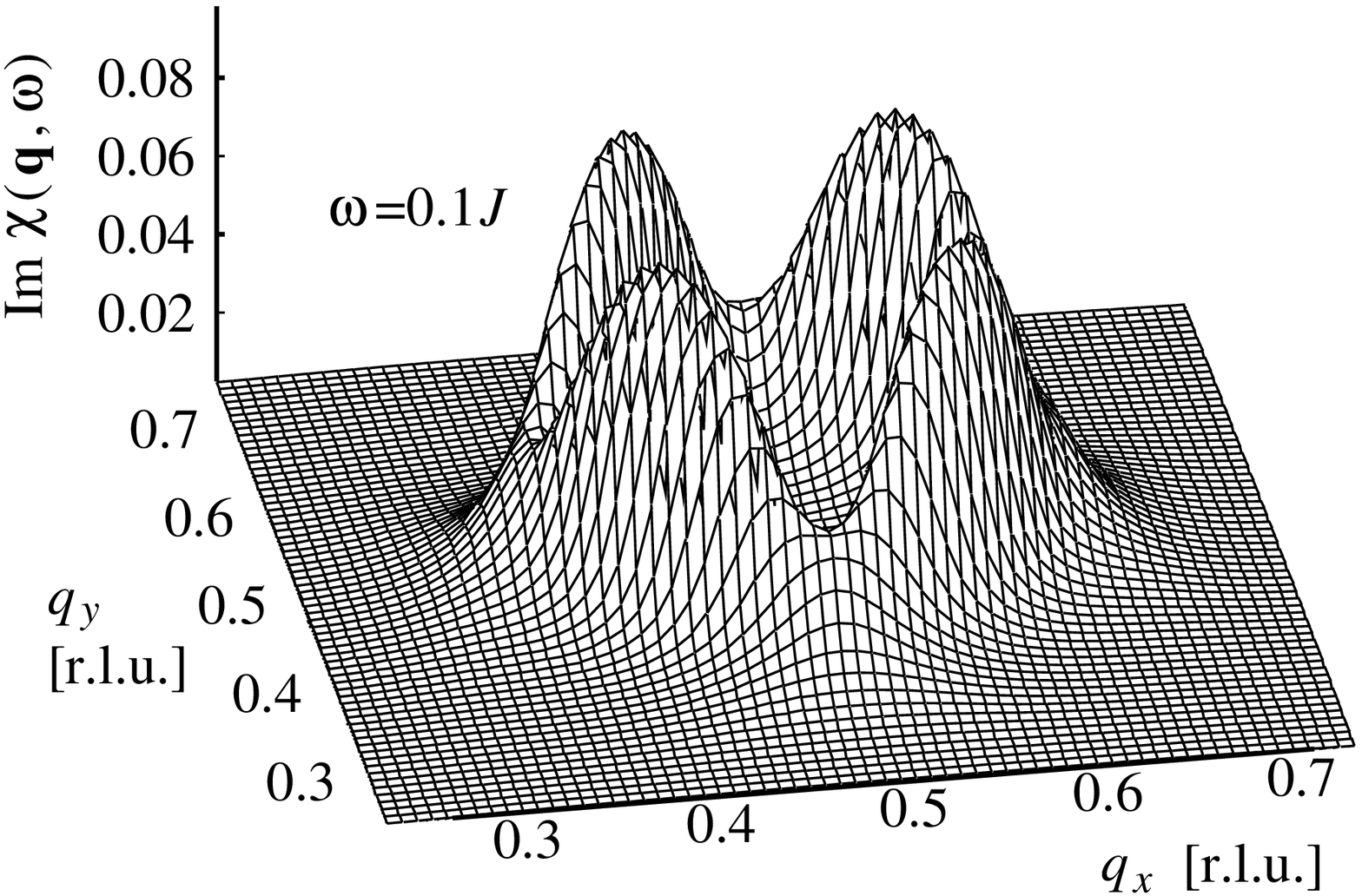} } \\
\end{center}
%
%
 %
%
%
\caption[\ ]{
  Wave-vector scan of $\chi''({\bf q},\omega)$ (in arb.\ units)
  around ${\bf q}=(\pi,\pi)$ in the SC state at
  $T\to 0$\,. $q_x, q_y$ are in units of $2\pi=$1\,r.l.u. The hole
  filling is $x=0.12$\,, energies are
  $\omega=0.51J=\omega_0$\,,
  $0.3J=0.7\omega_0$\,, $0.1J$ (from top to bottom). A quasi-particle damping
  $\Gamma=0.02J$ has been used.
  }
 \label{fig-incomm}
\end{figure}

We now turn to the discussion of the magnetic response in wave-vector
space. The resonance at 41\,meV, as well as its relative in underdoped
samples at energies $\omega_0 < 41$\,meV\,, is characterized as
a single (commensurate) peak at ${\bf q}=(\pi,\pi)=(1/2,1/2)$\,r.l.u.\
\cite{moo93,fon95,bou96,fon97}\,. Recent INS experiments by Mook
et.al.\ \cite{mook98}\,,
performed at an energy $\omega_i=24\mbox{\,meV}\approx 0.7 \omega_0$
on an Y\,Ba$_2$Cu$_3$O$_{6.6}$ sample with $\omega_0=34$\,meV gave evidence
for an incommensurate response.
It is dominated by four peaks, horizontally displaced from
$(\pi,\pi)$\,. The maximum intensity is strongly reduced
compared to $\omega=\omega_0$\,. Fig.\ \ref{fig-incomm} shows
$\chi''({\bf q},\omega)$ for $x=0.12$ at $T\to 0$ from the renormalized RPA
Eq.(\ref{eqn-rpa}) with vertex corrections omitted. At the energy
$\omega_0=0.51J$ (top panel) the susceptibility is commensurate
\cite{liu95}\,. When the energy is lowered to $\omega_i=0.7\omega_0=
0.3J$\,, the intensity drops dramatically, and four peaks appear at ${\bf
q}=(\pi\pm\delta,\pi)$ and $(\pi,\pi\pm\delta)$\,, as is seen in
the experiment \cite{mook98}\,. The
amount $\delta=0.1$\,r.l.u.\ of the displacement fits well the
observed value $0.105$\,r.l.u. The incommensurate
pattern may be characterized by the intensity ratio $I_h / I_d$\,, which is
the maximum intensity $I_h$ found in a horizontal scan through $(\pi,\pi)$\,,
related to the maximum $I_d$ from a diagonal scan. The numerical calculation
yields $I_h / I_d = 1.4$\,; from the data given in Ref.\
\cite{mook98} we estimate an experimental value $\lesssim 2.0$\,.
When the energy is further reduced, $I_h / I_d$ changes continuously
to values $<1$\,, and for $\omega<<\omega_0$ (bottom panel) four peaks
appear at ${\bf q}=(\pi\pm\delta',\pi\pm\delta')$
\cite{lu92,tan94}\,. At these low energies no information on
the {\bf q}-space structure has yet been obtained from INS, due to the
very dim intensity.

\begin{figure}[htb]
  \mysize \hsize
  \mygapsize 0.1\hsize
  \def\epsfsize#1#2{\epsfxsize}
  \epsfxsize=\mysize
  \parbox{\mysize}{ \epsffile{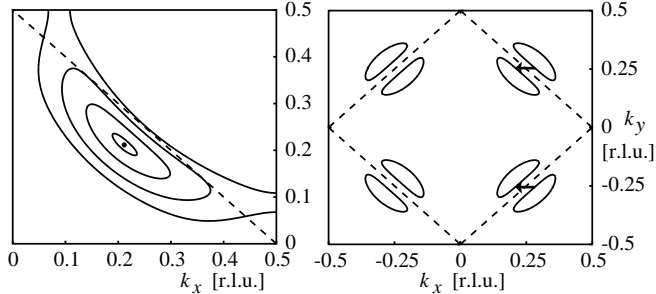} }
  \vspace*{-\mygapsize}
%
%
\vskip 30pt
\caption[\ ]{
  {\bf Left:} contour plot of $E({\bf k})$ in the SC state for
  $x=0.12$\,. Shown is the upper right 1/4 Brillouin zone
  (BZ) with $k_{x/y}$ in units of $2\pi=$1\,r.l.u. The node $E({\bf
  k})=0$ appears as a dot, encircled by contours $E({\bf k})=\omega/2$
  for $\omega=0.1J$ (most inner curve)\,, $0.3J$\,, $0.51J$\,, $0.7J$\,.
  {\bf Right:} contour for $\omega=0.3J$ and its by $(\pi,\pi)$  shifted
  image in the full BZ\,. Arrows indicate the best nesting vector relative to
  $(\pi,\pi)$\,.
  }
 \label{fig-cont}
\end{figure}

An explanation for the incommensurate pattern at $\omega=\omega_i$ can
be found in the quasi-particle dispersion $E({\bf k})$
of the d-wave SC state. Consider again the ph-excitation energies
$\Omega({\bf q}, {\bf k})$\,.
At low energy $\omega\to 0$ only excitations
between the nodes $E({\bf k})\gtrsim 0$\,, $E({\bf k} + {\bf
q})\gtrsim 0$ contribute, with
${\bf q}=(\pi\pm\delta',\pi\pm\delta')$
near $(\pi,\pi)$ \cite{lu92}\,. For small $\omega=0.1J$ this is still
the dominant process, see Fig.\ \ref{fig-cont} (left). The
situation changes for $\omega=\omega_i=0.3J$\,. The
contour $E({\bf k})=\omega_i/2$ contains a flat piece, which allows
for a nesting
contribution from (nearly) degenerate excitations
$E({\bf k})\approx E({\bf k}+{\bf q})=\omega_i/2$\,.
Following Ref.\ \cite{schulz90} the best nesting vector
is a horizontal (or vertical) offset to $(\pi,\pi)$, i.e., ${\bf
q}=(\pi\pm\delta,\pi)$ and $(\pi,\pi\pm\delta)$\,. This is illustrated
in Fig.\ \ref{fig-cont} (right), the offset is given by
$\delta = \arcsin[(\omega + \mu_f) / -\widetilde{t}]$\,.
The above reasoning has been given without consideration of the renormalized
RPA, Eq.(\ref{eqn-rpa}), since in fact $\chi\approx\chi^0$ for
energies below $\omega_0$\,. On the other hand, this is not the case
near the resonance energy $\omega_0=0.51J$\,, where $\chi$ is determined by
a pole in Eq.(\ref{eqn-rpa})\,, which produces the strong commensurate
peak displayed in the top panel of Fig.\ \ref{fig-incomm}\,.

In conclusion we find that the slave-boson mean-field theory, enhanced
by the renormalized RPA, gives a reasonable explanation for the INS
experiments on YBa$_2$Cu$_3$O$_{6+y}$ compounds. The energy and
vanishing damping of the ``41\,meV resonance'' is associated with the
threshold of ph-excitations at ${\bf q}=(\pi,\pi)$\,. The evolution of
the resonance with hole filling is accounted for, but the
damping is quite sensitive to the bandstructure (i.e., $t'\ne 0$). The
same theory explains the incommensurate structure at lower energies
as a dynamic nesting effect, specific to the d-wave SC state.

This work is supported by NSF under the \mbox{MRSEC} program
\mbox{DMR\,98-08941}\,. JB acknowledges financial support from
the Deutsche Forschungsgemeinschaft, Germany.

\sloppy
\bibliographystyle{prsty}

\end{document}